\documentstyle[12pt]{article}

\newcommand{\beq}{\begin{equation}}
\newcommand{\eeq}{\end{equation}}
\newcommand{\beqa}{\begin{eqnarray}}
\newcommand{\eeqa}{\end{eqnarray}}

\newcommand{\eq}[1]{(\ref{#1})}

\newcommand{\ra}{\rightarrow}
\newcommand{\del}{\partial}
\newcommand{\psib}{{\bar \psi}}
\newcommand{\teta}{\Theta_2(s{2\mu\over\beta},is{4\pi\over\beta^2})}
\newcommand{\integ}{\int_0^\infty}
\newcommand{\shR}{\mbox{sh}(Rs/4)}
\newcommand{\rinv}{{2\over R}}
\newcommand{\NP}[1]{ {\it Nucl.~Phys.} {\bf #1}}
\newcommand{\PL}[1]{ {\it Phys.~Lett.} {\bf #1}}
\newcommand{\Prep}[1]{ {\it Phys.~Rep.} {\bf #1}}
\newcommand{\PR}[1]{ {\it Phys.~Rev.} {\bf #1}}

\newcommand{\PTP}[1]{ {\it Prog.~Theor.~Phys.} {\bf #1}}

\newcommand{\MPL}[1]{ {\it Mod.~Phys.~Lett.} {\bf #1}}

\begin{document}
\topmargin 0pt
\oddsidemargin 1mm
\begin{titlepage}
\begin{flushright}
 LMU-TPW 94-18\\
 OU-HET 205 \\
 hep-th 9412149 \\
\end{flushright}
\setcounter{page}{0}
\vspace{15mm}
\begin{center}
{\Large Phase Diagram of Gross-Neveu Model at Finite Temperature,
        Density and Constant Curvature}
\vspace{20mm}

{\large Shinya Kanemura
\footnote{e-mail: kanemu@phys.wani.osaka-u.ac-jp}}\\
{\em Department of Physics, Osaka University\\
Machikaneyama 1-16, Toyonaka 560, Japan}\\
and\\
{\large Haru-Tada Sato
\footnote{JSPS Fellow, on leave of absence from Osaka Univ.\\
\phantom{mil}e-mail: hsato@lswes8.ls-wess.physik.uni-muenchen.de}}\\
{\em Sektion Physik der Universit{\"a}t M{\"u}nchen \\
Theresienstrasse 37, D-80333 M{\"u}nchen, Germany}\\
\end{center}
\vspace{7mm}

\begin{abstract}
We discuss a phase structure of chiral symmetry breaking in the
Gross-Neveu model at finite temperature, density and constant
curvature. The effective potential is evaluated in the leading
order of the $1/N$-expansion and in a weak curvature approximation.
The third order critical line is found on the critical surface in
the parameter space of temperature, chemical potential and constant
curvature.
\end{abstract}

\vspace{1cm}

\end{titlepage}
\newpage
\renewcommand{\thefootnote}{\arabic{footnote}}
\indent

The application of finite temperature field theories is important for
the restoration of electoweak symmetry and composite Higgs models, the
deconfinement phase of the quark-gluon plasma and hadronic physics.
The underlying concept of these approaches is a dynamical symmetry
breaking and the paradigm of this notion is digested in the chiral
Gross-Neveu models \cite{GN}, which show the properties of renormalizability,
asymptotic freedom and dimensional transmutation. The Finite temperature
analyses of the Gross-Neveu models were made two decades ago \cite{Tc}.
The mass spectrum is given \cite{spec} and the kink-antikink
configurations are not negligible in first order phase transitions
\cite{kink1,latt,kink2}. In order to give an analytical insight into
the finite temperature models and to compare with the results already
obtained in various cases \cite{Tc,HY}, we however concentrate only on
single particle state's configuration in the large $N$ limit approach.

The chiral phase transitions caused by an external curvature are discussed
in some two-dimensional models \cite{Ito,BK,SW}. In this paper, we would
like to examine the effective potential of the two-dimensional discrete
chiral Gross-Neveu model under both finite temperature and density on the
constant curvature spacetime and show its phase diagram in these external
parameters' space. We first reformulate the effective
potential for the Lagrangian
\beq
{\cal L}=\psib i\gamma^\mu\nabla_\mu\psi-{N\over2\lambda}\sigma^2
        -\sigma\psib\psi,                         \label{eq1}
\eeq
where $\psi$ is a $N$-component (or -flavor) fermion field and the
summation over the flavors is implicit in the fermion bilinear forms.
In broken phases, the dynamical mass is generated by a vacuum expectation
value of the auxiliary field $\sigma$ and the phase sturucure is
governed by the effective potential \cite{BK}
\beq
V(\sigma;R)={1\over2\lambda}\sigma^2
           -i{\rm tr}\int_0^\sigma d\sigma S(x,x)        \label{eq2}
\eeq
in which $S(x,y)$ is estimated in the proper time method and
\beq
S(x,x)=-i\integ ds e^{-s(\sigma^2+R/4)}{\sigma\over4\pi s}
        {Rs/4\over sh(Rs/4)}.                      \label{eq3}
\eeq
With the trace formula in the Euclidean spacetime
\beq
1=(4\pi s)^{D/2}{\rm Tr}\,\, exp(-s\del^2),                  \label{eq4}
\eeq
we can rewrite \eq{eq3} as an integral on the momentum trace so as to
estimate the contribution of momenta separately from curvature's effect
\beq
S(x,x)=-i\int_{-\infty}^{\infty}{d^2k\over(2\pi)^2}\integ ds\sigma
       {Rs/4\over\shR}e^{-s(k^2+\sigma^2+R/4)}.        \label{eq5}
\eeq
We then obtain a momentum-integral representation for the effective
potential \eq{eq2}
\beq
V(\sigma;R)={1\over2\lambda}\sigma^2+\int {d^2k\over(2\pi)^2}\integ
            {ds\over s}{Rs/4\over\shR}
            e^{-s(k^2+R/4)}(e^{-s\sigma^2}-1).        \label{eq6}
\eeq
If we take into account of a mass term, just replace $k^2\ra k^2+m^2$. As
for an extermal gauge field, we may generalize \eq{eq6} following the method
of ref.\cite{proper} as well. This expression is also convenient to
introduce the temperature as exhibited below. We here notice that the
limit $R\ra0$ is now easy to obtain the usual momentum-integral
representation
\beq
V(\sigma)={1\over2\lambda}\sigma^2+\int {d^2k\over(2\pi)^2}
          \ln({k^2+\sigma^2\over k^2}).                 \label{eq7}
\eeq

Before proceeding to a finite temperature model, we make some
remarks here. First, in order to incorporate the temperature, we must
break the general covariance and the curvature can not be defined in a
covariant manner accordingly. One possible way to incorporate curvature
seems to deal with the $(D-1)$-dimensional space curvature, however such
introduction of curvature in finite temperature systems becomes trivial
in two-dimensional models. Although the definition of curvature would be
temperature variant in the above sense, however we assume that the value
of the $R$ would be still a constant independently of the temperature. In
this consideration, we can employ the constant $R$ as an external parameter,
which is already defined in original covariant theories.
Second, as can be understood from the derivation of \eq{eq6}, the
introduction of temperature on the $k^2$ term in \eq{eq6} corresponds to
a kind of flat spacetime approximation, which is slightly different from
linear curvature expansions \cite{ELO}. We should note also that \eq{eq6}
can not pick up any contribution from winding modes.
There should be more controversies about this point as well as
kink-antikink mode, however, we would like to concentrate on \eq{eq6}
in order to separately observe its characteristic phase structure
how or whether the third order critical (tri-critical) point on the
$T$-$\mu$ plane \cite{Wolf} extends to the $R$-$T$-$\mu$ phase space.

We now introduce the temperature $T$ ($=1/\beta$) and the chemical
potential $\mu$ to \eq{eq6} in the usual way,
\beq
k^2\quad\ra\quad(\omega_n-i\mu)^2+k_1^2,\hskip 30pt
\int{dk_0\over2\pi}\quad\ra\quad{1\over\beta}\sum_n        \label{eq8}
\eeq
where $\omega_n=(2n+1)\pi/\beta$. Both integration over $k_1$ and summation
over $n$ can be easily performed and thus the bare potential for our finite
$R$-$T$-$\mu$ model becomes
\beq
V(\sigma;R,\beta,\mu)={1\over2\lambda}\sigma^2+{R\over4\beta}\integ
                      ds{\teta\over(4\pi s)^{1/2}\shR}e^{-s(R/4-\mu^2)}
                     (e^{-s\sigma^2}-1),                  \label{eq9}
\eeq
where $\Theta_2$ is the elliptic theta function. We can verify various
limits of this proper time representation at this stage. The limit of
both $\beta\ra\infty$ and $\mu\ra0$ coincides with \eq{eq6}. For the
finite $T$-$\mu$ effective potential we can obtain the following
expression taking $R\ra0$
\beq
V(\sigma;\beta,\mu)={1\over2\lambda}\sigma^2+{1\over\beta}\integ{ds\over s}
                    {\teta\over(4\pi s)^{1/2}}e^{s\mu^2}
                    (e^{-s\sigma^2}-1).                   \label{eq10}
\eeq
This is an alternative representation to ones already obtained in
\cite{Muta,Wolf}.
The limit $\beta\ra\infty$ with a finite value of $\mu$ turns out
\beq
V(\sigma;R,\mu)={1\over2\lambda}\sigma^2+{R\over4}\integ{ds\over4\pi s}
                {(1+2s\mu^2)^{-1/2} \over\shR}e^{-s(R/4-\mu^2)}
                (e^{-s\sigma^2}-1).                      \label{eq11}
\eeq
Although we should apply the renormalization condition
\beq
\lim_{T,\mu\ra0,R\ra R_0} {\del^2\over\del\sigma^2}
V(\sigma;R,T,\mu)\Bigr\vert_{\sigma=1} ={1\over\lambda_R},  \label{eq12}
\eeq
or equivalently,
\beq
{1\over\lambda}-{1\over\lambda_R}={R\over8\pi}\integ ds
  {1-2s\over\shR}e^{-s(1+R/4)}\Bigr\vert_{R\ra R_0},          \label{eq13}
\eeq
we however put $R_0=R$ in the following analysis. This gives same
counter term up to a finite renormalization as the result of Buchbinder-
Kirillova \cite{BK}, and consequently phase structure does not essentially
change. The renormalized effective potential of our model is therefore
\[
\hskip -55pt
V(\sigma;R,\beta,\mu)={1\over2\lambda_R}\sigma^2+{R\over16\beta}\integ
ds {e^{-sR/4}\over\shR}
\]
\beq
\hskip 35pt
\times[\,{1\over s}(e^{-s\sigma^2}-1){\sqrt{4\pi s}\over\beta}
e^{s\mu^2}\teta+\sigma^2e^{-s}(1-2s)\,].                   \label{eq14}
\eeq

Now we can write down the equations for the dynamical mass and the
critical surface for second order transition. The gap equation is
given by
\beq
0={1\over\lambda_R}+{R\over8\pi}\integ ds {e^{-sR/4}\over\shR}
[\,-e^{-s(\sigma^2-\mu^2)}{\sqrt{4\pi s}\over\beta}\teta +
                        e^{-s}(1-2s)\,].                   \label{eq15}
\eeq
Hereafter we adopt the value $\lambda_R=\pi$ for the renormalized coupling
constant of which value coincides with that of $T=\mu=R=0$ limit case.
The critical surface on which the second order phase transition occurs
is described by
\beq
\lim_{\sigma\ra0}{\del\over\del\sigma^2}
                 V(\sigma;R,\beta,\mu) =0,              \label{eq204}
\eeq
namely, by the following equation
\beq
0=\integ{ds\over4\pi s} [\, e^{-sR/4}{Rs/4\over\shR}
  \{ -e^{s\mu^2}{\sqrt{4\pi s}\over\beta}\teta
  +  e^{-s}(1-2s) \}  + 2se^{-s} \,].                     \label{eq16}
\eeq
We should note that this surface covers up the first order critical
surface in some region and becomes irrelevant to describe the second
order phase transition (for instance see Fig.1). This situation
is very similar to that of $T$-$\mu$ critical line \cite{Wolf}.

We can see that some known results are recovered again in particular limits.
First, let us consider the case of limit $R\ra0$. Each term in the above
integrand diverges in this case. Introducing a parameter $D$ to regularize
these integrals as
\beq
0=\integ{ds\over (4\pi s)^{D/2}}
[\, e^{-s} -e^{s\mu^2}{\sqrt{4\pi s}\over\beta}\teta \,],  \label{eq17}
\eeq
we obtain
\beq
\beta^{D-2}\Gamma(1-{D\over2})={2\over\sqrt{\pi}} (2\pi)^{D-2}
\Gamma({3-D\over2})\Re\zeta(3-D,{1\over2}+i{\beta\mu\over2\pi}),\label{eq18}
\eeq
where $\zeta$ is the generalized zeta function.
This is noting but the equation for the second order critical line on
the $T$-$\mu$ plane in the dimensional reguralization formalism \cite{Muta},
in which this equation is proved to reproduce Treml's result \cite{Trem}
in the limit $D\ra2$. Second, let us take the limit both $\beta\ra\infty$
and $\mu\ra0$ in Eq.\eq{eq16};
\beq
0= \integ{ds\over4\pi s} [\, e^{-sR/4}{Rs/4\over\shR}
  \{ -1 +  e^{-s}(1-2s) \}  + 2se^{-s} \,].                 \label{eq19}
\eeq
This becomes
\beq
    2 = \gamma + \psi(1+\rinv)+{4\over R}\psi'(1+\rinv),    \label{eq20}
\eeq
where $\gamma$ is the Euler constant and $\psi$ the digamma function.
Buchbinder and Kirillova's result is now recovered using asymptotic
expansion for $R\sim0$ \cite{BK};
\beq
           \gamma=\ln ({R\over2}).                   \label{eq201}
\eeq
The limit $\beta\ra\infty$ becomes the second order critical line on
$R$-$\mu$ plane (Fig.2);
\beq
0= \integ{ds\over4\pi s} [\, e^{-sR/4}{Rs/4\over\shR}
   \{ -{e^{s\mu}\over\sqrt{1+2s\mu^2}}
      + e^{-s}(1-2s) \} + 2se^{-s} \,],                 \label{eq21}
\eeq
which can be cast into the following simple form similarly to \eq{eq20}
\beq
 0 = 2 + \psi_d(\rinv;\rinv\mu^2)
       + {4\over R}\psi'_d(\rinv;\rinv\mu^2),            \label{eq22}
\eeq
where $\psi'_d$ means a derivative on $z$ of the function
\beq
\psi_d(z;a)\equiv
\integ ds{e^{-sz}-e^{sa}(1+2as)^{-1/2}\over e^s-1}.        \label{eq23}
\eeq

As in the literature \cite{Wolf} (see also \cite{Muta}), we observe
the tri-critical point on the $T$-$\mu$ plane. Our interest is to
examine how the tri-critical point extends to finite curvature
region as previously mentioned. In our computation, the determination of
the third order critical line owes to the following equation;
\beq
\lim_{\sigma\ra0}({\del\over\del\sigma^2})^2
                 V(\sigma;R,\beta,\mu) =0.              \label{eq24}
\eeq
The explicit expression (integral representation) for \eq{eq24} is
\beq
       0= \integ ds {se^{s(\mu^2-R/4)}\over\shR}
                    {\sqrt{4\pi s}\over\beta}\teta.      \label{eq25}
\eeq
The intersection between this surface \eq{eq25} and the surface \eq{eq16}
is just the tri-critical line which cuts away an irrelevant piece from
the surface \eq{eq16} and defines the division between the first and the
second order phase transitions. Fig.1 shows the determination of a
tri-critical point at $\mu=0.65$. The tri-critical point ${\bf C}$
is found from crossing between two lines ${\bf A}$ and ${\bf B}$ which
correspond to \eq{eq16} and \eq{eq25} respectively.

The surface \eq{eq25} reduces the following equation of line on the
$R$-$\mu$ plane
\beq
    0=\integ ds {se^{sz}\over e^s-1}(1+2sz)^{-1/2},
            \hskip 20pt z=\rinv\mu^2,                    \label{eq28}
\eeq
however this line does not intersect with the second order critical line
\eq{eq22} in the region $\mu\leq1$ on the $R$-$\mu$ plane. There is thus
no tri-critical point in that region. In Fig.2, we draw the second order
critical line \eq{eq22}. It monotonically extends to the point
$(\mu,R)=(1,3.84)$, which we abbreviated in the figure.

Now we entirely explain how we draw the phase diagram of our model. First,
we start from a well-known point on the $T$-$\mu$ plane. The tri-critical
point can be found from \eq{eq17} and the surface \eq{eq25} with
the limit $R\ra0$
\beq
0=\integ ds e^{s\mu^2}{\sqrt{4\pi s}\over\beta}\teta,     \label{eq26}
\eeq
which coincides with the equation \cite{Wolf}
\beq
       \Re \zeta(3,{1\over2}+i{\beta\mu\over2\pi})=0.     \label{eq27}
\eeq
For finite curvature case, we similarly repeat this kind of analysis
using \eq{eq16} and \eq{eq25} to obtain the third order critical line.
Next, with the aid of \eq{eq16}, we can draw the second order critical
surface. Finally, we only have to determine the surface where the first
order phase transition occurs. Varying the parameter $T$ for fixed $R$
and $\mu$, we carefully observe the shapes of the effective potential
\eq{eq14} until the symmetry restores.

The results are summarized in Fig.3. The Critical lines at $\mu=0,0.5,0.6,
0.65,0.7,0.8,1.0$ are depicted. Increasing the value of $\mu$ from zero,
we firstly find one tri-critical point at $(T,R)=(0.25,0.7)$ when
$\mu=0.585$. After then, we observe two tri-critical points at both edges
of the first order critical line (for example, see $\mu=0.6$ case in Fig.3)
until $\mu=0.608$. When $\mu=0.608$, we obtain two tri-critical points
$(0.21,1.1)$ as well as $(0.32,0)$ (the latter point is well-known result).
For the interval after $\mu=0.608$ to $\mu=1$, we have only one tri-critical
point for each value of $\mu$. The tri-critical line would penetrate
$R-\mu$ plane in the region $1<\mu$.

We have observed the behaviour of the effective potential in the large
$N$ leading order as well as in a flat spacetime approximation and have
showed the third order critical line in the $R$-$T$-$\mu$ space.
In spite of these approximations, we have obtained a smooth interpolation
between critical values of temperatures and curvatures. For example,
when $\mu=0$, the point $T_c=0.57$ is smoothly connected to the point
$R_c=2.6$. As a result of this feature, the tri-critical line is extended
into finite $R$ region smoothly.

In higher dimensional Gross-Neveu models, the chiral symmetry
breaking is of second order ($2 \leq D <4$) in the large $N$ approach
\cite{Muta,3d} while the first order transition ($D=3,4$) appears in the
presence of curvature \cite{ELO}. Also in a lattice calculation
(without curvature), the existence of first order transition is
reported in three dimensions \cite{3dlat}. It would be interesting
whether the dimensional extension of our model would give an analytical
or quantitative connection between these kind of transitions in 3 or 4
dimensions. The Massive Gross-Neveu model is another interest of our
approach. The addition of a slight mass term does not entirely change the
phase structure of the finite $T$-$\mu$ model excepting that the second order
transition disappears \cite{mass}. We would like to pay attention to
what our model would have an effect on this point.

\vspace{1cm}
\noindent
{\em Acknowledgments}

The authors would like to thank T. Inagaki, T. Kouno, and T. Muta for
valuable conversations, J. Wess and H. Suzuki for useful suggestions.
\newpage

%
\pagebreak[4]
\topmargin 0pt
\oddsidemargin 5mm
\centerline{\bf FIGURE CAPTIONS}

\begin{description}
\item[{\bf Fig.1}:]
Determination of the tri-critical point {\bf C} at $\mu=0.65$.
The {\bf C} is determined from crossing of the
second order critical surface {\bf A} and the surface {\bf B}.
Dashed line {\bf D} is the first order critical line.
\item[{\bf Fig.2}:]
Phase structure on $R$-$\mu$ plane. {\bf A} and {\bf D} follow
the notation of Fig.1. Region {\bf S} is the symmetric phase and
{\bf B} the broken phase.
\item[{\bf Fig.3}:]
Critical lines at various values for $\mu$.
\end{description}
\end{document}